\documentclass[prd,aps,a4paper,superscriptaddress,twocolumn,nofootinbib]{revtex4} %
\usepackage[T1]{fontenc}
\usepackage{graphicx}
\usepackage{color}
\usepackage{dcolumn}
\usepackage{bm}
\usepackage{slashed}
\usepackage{amsmath}
\usepackage{latexsym}
\usepackage{amssymb}
\usepackage{mathrsfs}
\usepackage{amsfonts}
\usepackage{url}
\usepackage{float}
\usepackage[section]{placeins}
\allowdisplaybreaks
\usepackage{diagbox}
\usepackage{multirow}
\usepackage{makecell}
\usepackage{natbib}
\usepackage{xcolor}
\usepackage{comment}

\begin{document}
\title{Gravitational Wave Mixture Separation for Future Gravitational Wave Observatories Utilizing Deep Learning}
	
\author{CunLiang Ma}
\affiliation{School of Information Engineering, Jiangxi University of Science and Technology, Ganzhou, 341000, China}
	
\author{WeiGuang Zhou}
\affiliation{School of Information Engineering, Jiangxi University of Science and Technology, Ganzhou, 341000, China}
	
\author{Zhoujian Cao
\footnote{corresponding author}} \email[Zhoujian Cao: ]{zjcao@amt.ac.cn}
\affiliation{Institute of Applied Mathematics, Academy of Mathematics and Systems Science, Chinese Academy of Sciences, Beijing 100190, China}
\affiliation{School of Fundamental Physics and Mathematical Sciences, Hangzhou Institute for Advanced Study, UCAS, Hangzhou 310024, China}
	
\begin{abstract}
        Future GW observatories, such as the Einstein Telescope (ET), are expected to detect gravitational wave signals, some of which are likely to overlap with each other. This overlap may lead to misidentification as a single GW event, potentially biasing the estimated parameters of mixture GWs. In this paper, we adapt the concept of speech separation to address this issue by applying it to signal separation of overlapping GWs. We show that deep learning models can effectively separate overlapping GW signals. The proposed method may aid in eliminating biases in parameter estimation for such signals.
\end{abstract}

\maketitle

\section{Introduction\label{section:1}}

The field of gravitational-wave (GW) detection has witnessed remarkable progress since the first direct detection \cite{1,2,3,4,5,6,7}. The third observing run (O3) of GW detection ended in spring 2020, boosting the total number of confident events to above 90, with an event rate currently standing at 1.5 per week \cite{8,9,10,11}. However, the upcoming third-generation (3G) detectors such as Einstein Telescope (ET) \cite{12,13} and Cosmic Explorer (CE) \cite{14,15}, envisioned in the 2030s, promise a significant leap forward. This enables the detection rate of above $10^{5}$ per year at cosmological distances. The surge in detection rate, along with the remarkable enhancement of sensitivity across both lower and higher frequency ranges in 3G detectors, will significantly extend the duration of signals within the sensitivity band. As a consequence, the probability of GW signals overlapping in these 3G detectors will become significant \cite{16}, posing potential challenges for the GW search and parameter estimation.

As early as 2009, T. Regimbau and Scott A. Hughes delved into the effects of binary inspiral confusion on the sensitivity of ground-based GW detectors \cite{16}. They emphasized the necessity for rigorous data analysis to disentangle mixture signals. Since then, numerous studies have focused on analyzing the strain with mixture signals. Y. Himemoto \textit{et al.}, utilizing Fisher matrix analysis, explored the statistical ramifications of mixture GWs on parameter estimation \cite{17}. Their findings revealed that mixture signals can introduce notable statistical errors or systematic biases, especially when the coalescence times and redshifted chirp masses of the mixture GWs are closely matched. A realistic distribution analysis further indicated that mergers occurring within a second of each other are common occurrences over a year in 3G detectors \cite{18}.

Modern data analysis techniques for parameter estimation typically assume the presence of a single signal amidst background noise. However, when two or more GWs are simultaneously detected, their signals overlap, creating a distorted, non-physical waveform. This leads the sampling software to identify parameter sets aligned with this composite waveform, rather than the individual signals \cite{19}. Experimental results from P. Relton \textit{et al.} demonstrated that, in most instances, current parameter estimation methods can accurately assess the parameters of one of the mixture events \cite{19}. Notably, if one signal is at least three times stronger than the other, the louder signal's source parameters remain unaffected \cite{19}. By applying a narrow prior on the coalescence time, obtained during the GW detection phase, it may be feasible to accurately recover both posterior parameter distributions \cite{20}. Experiments conducted by E. Pizzati \textit{et al.} showed that parameter inference remains robust as long as the coalescence time difference in the detector frame exceeds 1 second \cite{20}. Conversely, when this time difference is less than 0.5 seconds, significant biases in parameter inference are likely to emerge \cite{20}. Upon comparing the effects of mixture signals on coefficients at various post-Newtonian (PN) orders, it has been determined that, overall, the 1PN coefficient experiences the greatest impact. The findings further indicate that, although a significant proportion of mixture signals introduce biases in PN coefficients, which individually might suggest deviations from General Relativity (GR), collectively, these deviations occur in random directions. As a result, a statistical aggregation of these effects would still tend to align with GR \cite{21}. Quantifying source confusion within a realistic neutron star binary population reveals that parameter uncertainty generally rises by less than 1\%, except in cases where overlapping signals exist with a detector-frame chirp mass difference of \( \lesssim 0.01 \, M_\odot \) and an overlap frequency of \( \gtrsim 40 \, \text{Hz} \) \cite{22}. Among \(1 \times 10^6\) simulated signals, only 0.14\% fall within this specific range of detector-frame chirp mass differences, yet their overlap frequencies are usually below 40 Hz \cite{22}.

Apart from the task of parameter estimation, several studies focus on exploring the impact of overlapping signals on gravitational wave detection. Within the CWB framework for GW searching, most signals resulting from closely merged events will only be detected as a single trigger \cite{23}. In the context of the PyCBC framework and the search for binary black hole (BBH) events, it has been noted that when the relative merger time exceeds 1 second, the search efficiency diminishes by approximately 1\% \cite{23}. In cases where the relative merger time is less than 1 second, the search efficiency drops by 26\% because most paired signals are either detected by a single trigger or not detected at all \cite{23}. The biases in the estimation of the PSD will negatively impact the sensitivity of the 3G ground-based GW detectors, especially considering the large population of overlapping signals \cite{24}. The confusion noise's contribution to the signal-to-noise ratio (SNR) is considerably lesser than that of the instrumental noise \cite{24}.

Certain studies focus on refining data processing techniques to address the challenge posed by overlapping signals. J. Janquart \textit{et al.} analyze the overlapping binary black hole merger with hierarchical subtraction and joint parameter estimation \cite{25}. They find that joint parameter estimation is usually more precise but comes with higher computational costs. J. Langendorff \textit{et al.} first utilize normalizing flows for the parameter estimation of overlapping GW signals \cite{26}. Compared to the traditional Bayesian method, the normalizing flow results in broader posterior distributions, whereas the Bayesian-based approach tends to become overconfident, potentially overlooking the injection \cite{26}.

Recently, we have proposed a novel framework (MSNRnet) aimed at accelerating the matched filtering process for GW detection \cite{27}. This is achieved by incorporating deep learning techniques for waveform extraction and discrimination. However, as the waveform extraction stage solely captures one waveform, in scenarios where multiple signals overlap, there is a possibility that the MSNRnet framework may overlook one of the overlapping signals.

Real-world speech communication frequently takes place in vibrant, multi-speaker settings \cite{28}. To function effectively in these environments, a speech processing system must possess the capability to distinguish and separate speeches from various speakers. While this endeavor comes naturally to humans, it has been exceedingly challenging to replicate in machines. However, in recent years, deep learning strategies have notably pushed the boundaries of this problem \cite{29,30,31,32,33,34,35,36,37,38,39,40}, surpassing traditional techniques like independent component analysis (ICA) \cite{41} and semi-nonnegative matrix factorization (semi-NMF) \cite{42}. The mixed speech can be compared to mixed GW signals. Drawing inspiration from the task of speech separation, this study marks the first attempt to apply deep learning to GW separation. The proposed method for GW signal separation holds potential for future applications in GW search and parameter estimation. Furthermore, this work serves as a complement to the existing tasks of deep learning applied to GW data processing, including end-to-end GW signal search \cite{43,44,45,46,47,48,49,50,51,52,53,54,55,56,57,58,59,60,61}, parameter estimation \cite{62,63,64,65,66}, waveform or envelope extraction \cite{67,68,69,70}, GW source localization \cite{71,72,73,74}, and glitch classification \cite{75,76,77,78}. Since the GW components buried in noise, the GW separation task is more challenging than speech separation.

In this work, we first explored the potential of utilizing deep learning for GW separation. We find that the mixture strain with noise and multi-signals can be separated. 

\section{METHOD FOR GW SEPARATION\label{section:2}}

In the early stages of applying deep learning to speech separation, the preprocessing phase typically involved converting mixed sound into a time-frequency representation \cite{79,80,81,82}, isolating source bins via time-frequency masks, and synthesizing waveforms via invert time-frequency transform. However, challenges arose, including the optimality of Fourier decomposition and the need to handle both magnitude and phase in the complex STFT domain. This often led to methods that only adjusted the magnitude, ultimately limiting separation performance. In 2018, Luo \textit{et al.} introduced the Time-domain Audio Separation Network (TasNet) \cite{28}. This neural network was designed to directly model the time-domain mixture waveform through an encoder-separation-decoder framework, where the actual separation occurred at the encoder's output. The following year, they further refined TasNet, evolving it into Conv-TasNet \cite{29}. The key innovation of Conv-TasNet was the use of a Temporal Convolutional Network (TCN) for the separation component, consisting of stacked one-dimensional dilated convolutional blocks. In 2020, the same team proposed DPRNN \cite{32}, which incorporated a dual-path RNN for the separation phase. Later that year, J. Chen \textit{et al.} enhanced DPRNN, giving birth to DPTNet \cite{34}. This advancement replaced the dual-path RNN module with a dual-path transformer module. We have utilized all three iterations of TasNet—Conv-TasNet, DPRNN, and DPTNet—for the task of GW separation. Among these, we find that DPRNN has proven to be superior to the other two methods. So, in this work, we focus on DPRNN for GW separation.

Suppose that the strain captured by the interferometer, denoted as \(d(t)\), can be regarded as a combination of a noise component, \(n(t)\), and the GW component, \(h(t)\).

\begin{equation}
        d\left(t\right) = n\left(t\right) + h\left(t\right)
\end{equation}        

\noindent For the GW separation task, the GW component comprises multiple signals, represented as $h\left(t\right)=\sum_{i=1}^{N}{h_i(t)}$, where \(h_i(t)\) signifies each individual GW signal, and \(N\) signifies the overall count of GW signals existing in the analyzed data segment. In this work, we will solely focus on the scenario where there are two signals \(h_A(t)\) and \(h_B(t)\) present in the data. So

\begin{equation}
        d\left(t\right) = n\left(t\right) + h_A\left(t\right) + h_B\left(t\right).
\end{equation}

\begin{figure*}[htbp]
\includegraphics[scale=0.46]{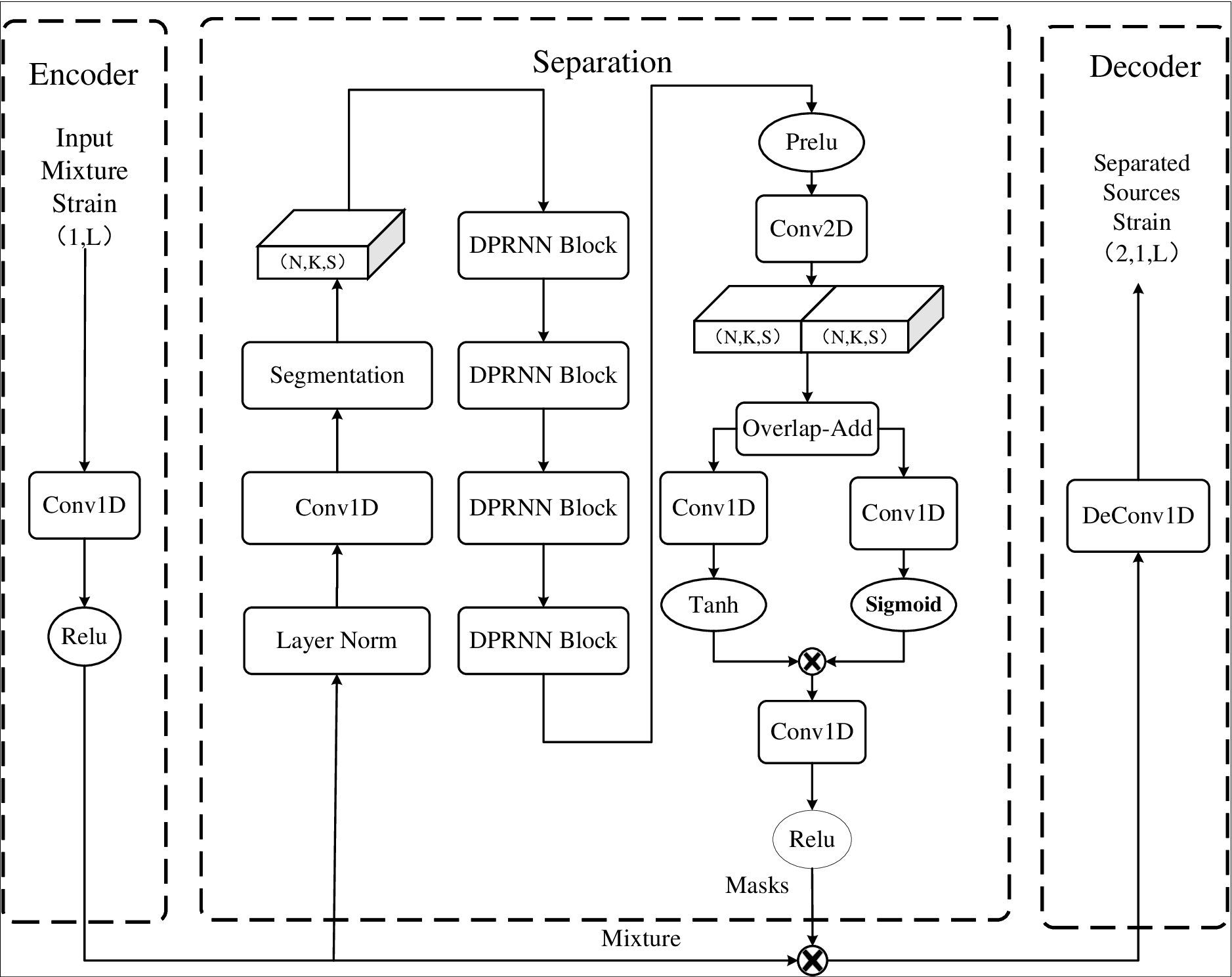}
\caption{\label{fig:1}Schematic diagram of the GW separation model. The whole process can be divided into Encoder, Separation, and Decoder stages.}
\end{figure*}

\noindent We aim to directly estimate $h_A\left(t\right)$ and $h_B\left(t\right)$ from $d\left(t\right)$. The TasNet-like framework decomposes the signal separation task into three stages: Encoder, Separation, and Decoder, and the overall framework for GW separation is shown in Fig.~\ref{fig:1}. During the Encoder stage, the input signal is encoded into a hidden layer feature $F$. In the Separation stage, masks ($M_A$ and $M_B$) for each signal component are evaluated. Subsequently, the Decoder stage utilizes these masked features to obtain the separated output as follows:

\begin{align}
        \widetilde{h}_A &= \text{Decoder}\left(M_A \odot F\right), \\
        \widetilde{h}_B &= \text{Decoder}\left(M_B \odot F\right).
\end{align}
        
\noindent where $\odot$ denotes the Hadamard product. The Encoder, Separation, and Decoder stages can be likened to the STFT, time-frequency masking, and inverse STFT stages respectively, of signal separation utilized by the short-time Fourier transform. In the subsections that follow, we will elaborate on the three stages of GW separation.

\subsection{Encoder stage\label{subsection:2-1}}

Suppose the Encoder receives an input signal $s \in \mathbb{R}^{1 \times L}$, where $L$ denotes the number of time samples of the input strain. Through the Encoder stage, we get the signal feature $F \in \mathbb{R}^{C \times L}$ by

\begin{equation}
        F = \text{ReLU}\left(\text{Conv1D}\left(s\right)\right),
\end{equation}
        
\noindent where in the 1D convolutional layer, C=256 filters are used, and the filter size is configured to 2.

\subsection{Separation stage\label{subsection:2-2}}

The input of the separation stage is signal feature $F$ and the output generates two feature masks namely $M_A$ and $M_B$. The signal feature $F$ is initially passed through Layer Normalization and a Conv1D layer, undergoing transformation into a tensor representation having a shape of $\mathbb{R}^{N \times L}$ where $N=64$ represents the number of Conv1D filters. Afterward, the tensor sequentially undergoes a segmentation operation, followed by processing through four DPRNN blocks, and concludes with an overlap-add operation. In the segmentation step, the 2D tensor undergoes a transformation into a 3D tensor through sub-frame alternation. This transformed tensor is then relayed to a stack of DPRNN blocks, where both local and global modeling are alternately and interactively employed. Upon completion of DPRNN processing, the output from the final layer is conveyed to a 2D convolutional layer and subsequently reverted to two 2D tensors via the Overlap-Add operation. These tensors are then simultaneously processed through two distinct convolutional modules equipped with different activation functions: Tanh and Sigmoid. Following this, the tensors are combined and subjected to a ReLU activation function, ultimately yielding two masks, designated as $M_A$ and $M_B$.

\subsubsection{Segmentation and Overlap-Add\label{subsection:2-2-1}}

\begin{figure*}[htbp]
        \includegraphics[scale=0.42]{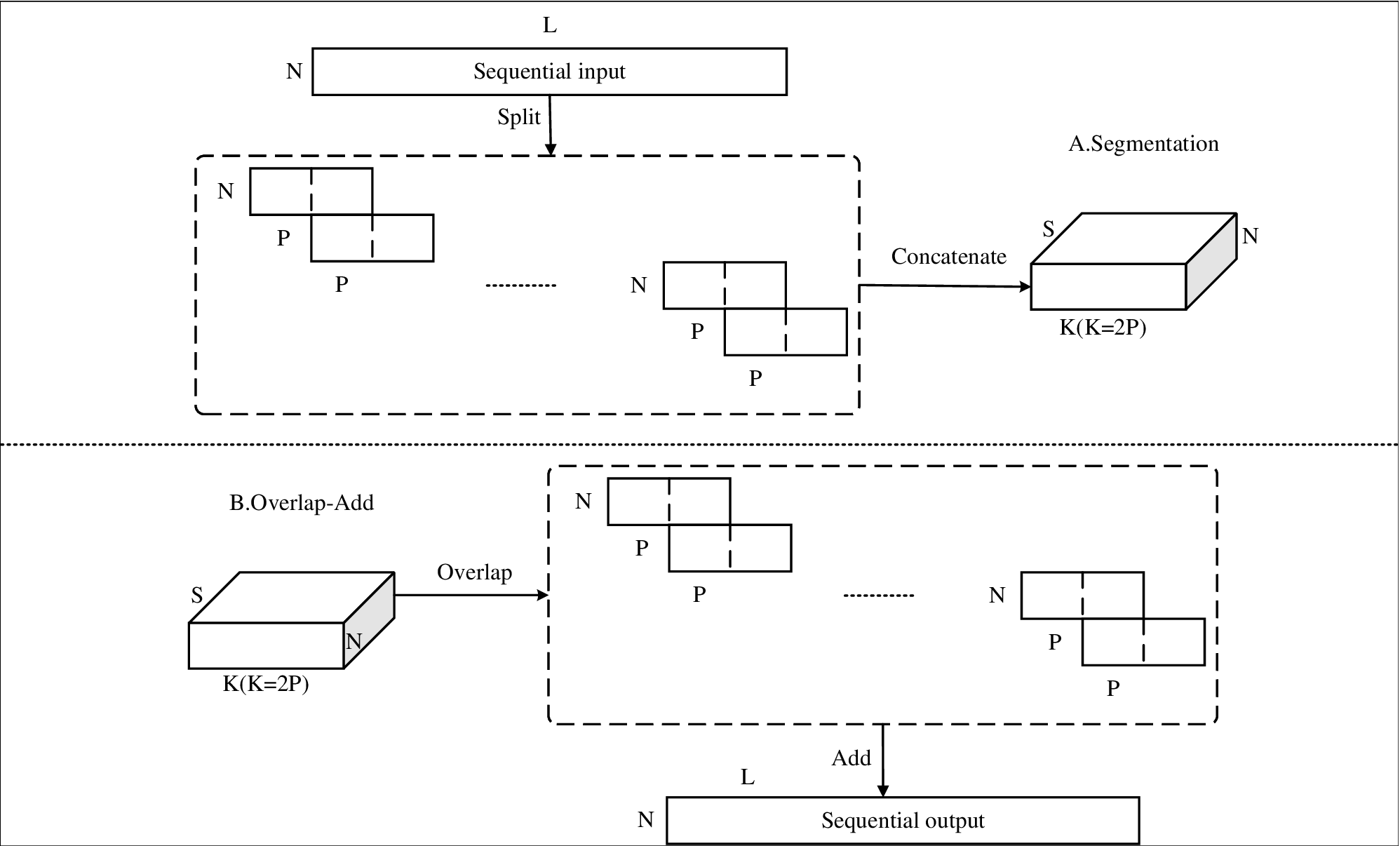}
        \caption{\label{fig:2}Schematic diagram of Segmentation process (A) and Overlap-Add process (B). The Segmentation process can transfer the 2D Tensor to the 3D Tensor, while the Overlap-Add process can transfer the 3D Tensor to the 2D Tensor.}
\end{figure*}

Fig.~\ref{fig:2} shows the flow chart of the Segmentation and Overlap-Add step in the separation stage. Let the input of the segmentation is a 2D tensor $F$ and the output of the segmentation is a 3D tensor $T$. For the segmentation stage, we first split the 2D tensor to $S$ small tensors ($D_i \in \mathbb{R}^{N \times K}$, $i \in \{1, 2, \ldots, S\}$). Then concatenate all the small 2D tensors together to form a 3D tensor $T = [D_1, D_2, \ldots, D_S] \in \mathbb{R}^{N \times K \times S}$. In this work $K = 250$ and $S = 134$.

Suppose the output of the last DPRNN block as $T_{B+1} \in \mathbb{R}^{N \times K \times S}$, then the Overlap-Add step can be seen as the inverse process of the Segmentation step. It applies the $S$ 2D tensors to form output $Q \in \mathbb{R}^{N \times L}$. Initially, we split the 3D tensor into $S$ 2D tensors and aligned according to real-time. Following this, we added the $S$ 2D tensors up and got one 2D tensor.

\subsubsection{DPRNN block\label{subsection:2-2-2}}

\begin{figure*}[htbp]
        \includegraphics[scale=0.53]{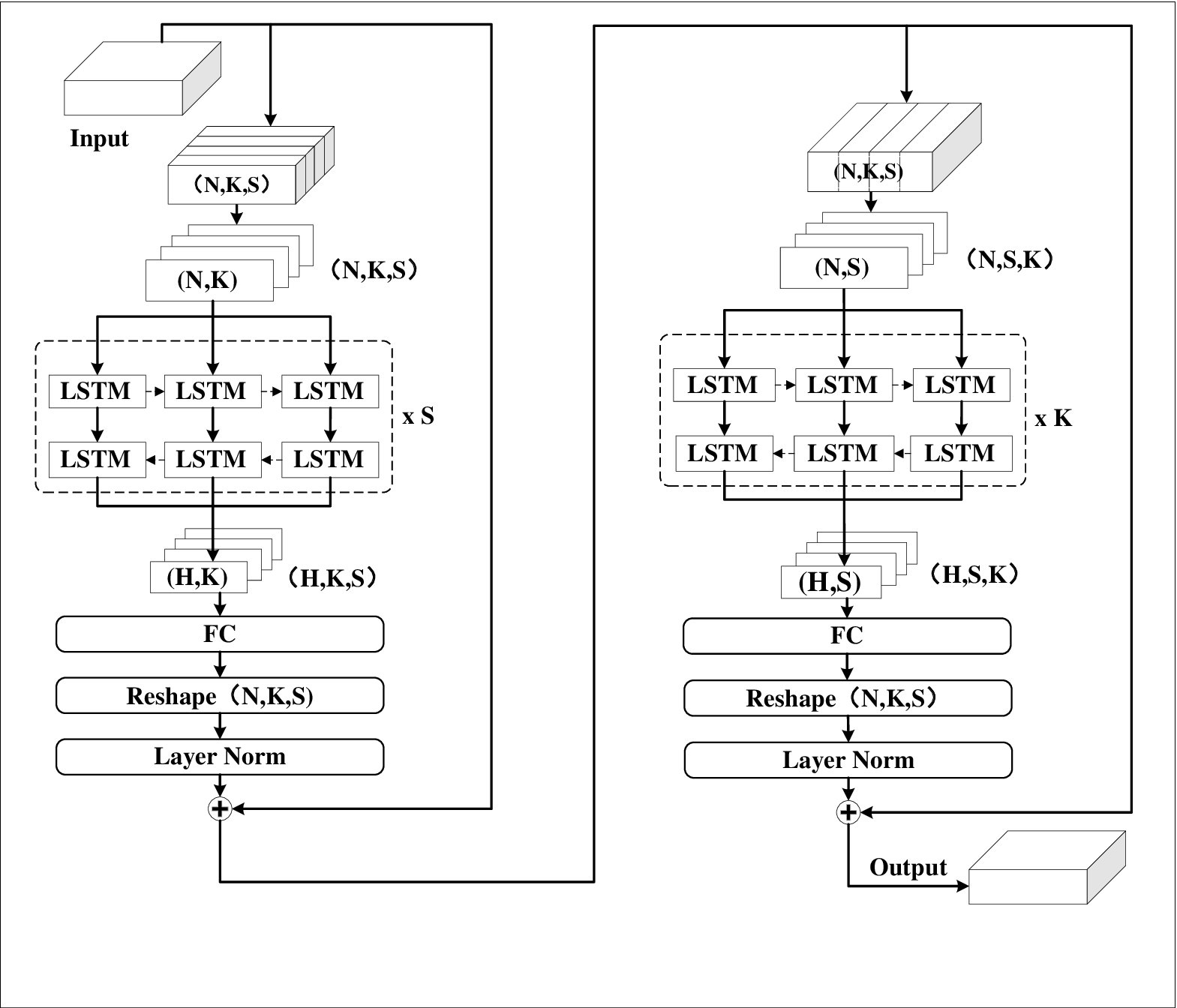}
        \caption{\label{fig:3}Schematic diagram of DPRNN block that maps a 3D tensor to another 3D tensor.}
\end{figure*}

The segmentation output $T$ is subsequently forwarded to a stack consisting of 4 DPRNN blocks. Each block maps a 3D tensor into another 3D tensor of the same shape. Let’s take the map $T_i \rightarrow T_{i+1}$ as an example to illustrate the calculation process of a DPRNN block. The flow chart depicting the DPRNN block is illustrated in Fig.~\ref{fig:3}. Initially, the input tensor is processed through a local modeling block, followed by a global modeling block. The key distinction between these two blocks lies in their approach to signal slicing. Specifically, the local modeling block slices the 3D tensor based on the third indicator, whereas the global modeling block performs slicing using the second indicator. For brevity, we only detail the mathematical expression pertaining to local modeling in this context.

Suppose the input of the local modeling is $T_i$ and the output is $\hat{T}_i$. We first put each divided chunk to a bidirectional LSTM block and concatenate them together to get a tensor $U_i \in \mathbb{R}^{H \times K \times S}$. In this work, we set $H$ to 256.

\begin{equation}
        U_i=\underset{j}{\operatorname*{Concatenate}}\text{ BiLSTM}(T_i[:,:,j]),
\end{equation}
      
\noindent where $T_i[:,:,j] \in \mathbb{R}^{N \times K}$ is the sequence defined by chunk $j$. We then apply a fully connected layer to the tensor $U_i$ and obtain $\widehat{U}_i \in \mathbb{R}^{N \times K \times S}$ as follows:

\begin{equation}
        \widehat{U}_i = \underset{j}{\operatorname*{Concatenate}\,}GU_i[:,:,j]
\end{equation}

\noindent where $G \in \mathbb{R}^{N \times H}$. Then Layer normalization is applied to $\widehat{U}_i$ as follows:

\begin{equation}
        LN(\widehat{U}_i)=\frac{\widehat{U}_i-\mu(\widehat{U}_i)}{\sqrt{\sigma(\widehat{U}_i)+\epsilon}}\odot z+r
\end{equation}

\noindent where $z, r \in \mathbb{R}^{N \times 1}$ are the rescaling factors, $\epsilon$ is a small positive number for numerical stability, and $\mu(\cdot)$ and $\sigma({\widehat{U}}_i)$ represent the mean and standard deviation operators, respectively. Then we get $\widehat{T}_i$ as follows:

\begin{equation}
        \widehat{T}_i = T_i + LN(\widehat{U}_i).
\end{equation}

\noindent Put the 3D tensor $\widehat{T}_i$ to the global modeling block, we then get the output of the DPRNN block $T_{i+1}$.

\subsection{Decoder stage\label{subsection:2-3}}

The Decoder stage maps the masked Encoded feature $F_{M_i} = M_i \odot F \in \mathbb{R}^{C \times L}$ to separated signal. Each element in $F_{M_i}$ (which can be likened to feature values) may be viewed as a component of a hidden vector (comparable to feature vectors) at a specific time.

\begin{equation}
        \tilde{h}_i=ConvTranspose1d(F_{M_i}),
\end{equation}

\noindent The hidden vectors can be regarded as the adjustable parameters of the transposed convolutional layer. This layer accepts $N$ input channels and outputs a single channel. Its purpose is to decrease the channel count of the masked encoded features from $C$ to $1$. By configuring the kernel size as $2$, stride as $1$, and padding as $0$, the transposed convolution preserves the length of the time series at $L$. As a result, the masked encoded features are reconstituted into a one-dimensional time series, denoted as $\widetilde{h}_i \in \mathbb{R}^{1 \times L}$.

\section{DATA FOR TRAINING AND TESTING\label{section:3}}

In this paper, we concentrate on the Einstein Telescope, which could potentially consist of three detectors arranged in a triangular configuration. For simplicity, we limit our analysis to just one of these detectors. We utilize the PyCBC package \cite{83,84,85,86} for synthesizing data, which aids in training, validation, and testing processes. The strain captured by the detector can be represented as a combination of noise and two mixture signals: $n(t) + h_A(t) + h_B(t)$, where $n(t)$ signifies the noise component. This noise is generated using the power spectrum density (PSD) linked to the Einstein Telescope, which offers insights into the detector's sensitivity at various frequencies. Specifically, we use EinsteinTelescopeP1600143 to simulate this noise.

Both \( h_A(t) \) and \( h_B(t) \) are generated through a linear combination of \( h_+(t) \) and \( h_\times(t) \), which are accurately modeled by SEOBNRv4. In our waveform simulation, the masses of the two black holes range from \( (10M_\odot, 80M_\odot) \). The dimensionless spin is randomly sampled within the interval \( (0, 0.998) \). Additionally, the declination and right ascension are uniformly sampled across the entire sphere. During the simulation of \( h_+(t) \) and \( h_\times(t) \), the luminosity distance from the astrophysical source to Earth is fixed at 4000 Mpc.

In the training phase, the amplitudes of \( h_A(t) \) and \( h_B(t) \) undergo random rescaling to align with two randomly generated signal-to-noise ratios (SNRs) falling between 5 and 20. Furthermore, the peak amplitude times of \( h_A(t) \) and \( h_B(t) \) are randomly positioned between 50\% and 95\% of the designated time window, which spans a duration of 4 seconds. The entire simulation operates at a sampling frequency of 4096 Hz.

\section{PERFORMANCE OF THE GW SEPARATION NETWORK\label{section:4}}

Previous studies examining data processing of overlapping gravitational wave (GW) strains have primarily focused on how GW overlapping affects traditional GW data processing methods, such as matched filtering for GW detection \cite{23} and Bayesian posterior sampling for parameter estimation \cite{20}. Recently, the normalizing flow has emerged as a new technique for parameter estimation of overlapping GW strains \cite{26, 84}. In our study, we propose the utilization of signal separation via deep learning for the analysis of overlapping GW strains.

The gravitational wave (GW) separation network can be considered a parameterized system. The network's output includes the waveforms of the estimated clean gravitational wave signals. To optimize the performance of the proposed model, we train it using utterance-level permutation invariant training (uPIT) \cite{88}, aiming to maximize the scale-invariant signal-to-noise ratio (SI-SNR) \cite{28}. SI-SNR is defined as:

\begin{align}
        s_{target} &= \frac{\langle \tilde{h}, h \rangle h}{\|h\|^2} \\
        e_{noise} &= \tilde{h} - s_{target} \\
        \text{SI-SNR} &:= 10 \log_{10} \frac{\|s_{target}\|^2}{\|e_{noise}\|^2}
\end{align}
    
\noindent where $\tilde{h} \in \mathbb{R}^{1 \times L}$ and ${h} \in \mathbb{R}^{1 \times L}$ are the estimated and target clean sources respectively, $L$ denotes the length of the signals, and $\tilde{h}$ and $h$ are both normalized to have zero-mean to ensure scale-invariance. During the training phase, the Adam method is used. A learning rate of $10^{-5}$ is established. The system undergoes 20 epochs of training. During the training stage, we assume that the peak time of signal A lags behind that of signal B. In other words, typically, signal A is only disrupted by the inspiral stage of signal B, whereas signal B experiences interference from the entire signal process, encompassing the inspiral, merger, and ringdown stages.

In this section, we explore the performance of the GW separation network. Prior researches \cite{19,20} have established that the accuracy of parameter estimation for the two sources can be notably influenced by both the peak time difference and the SNR difference. Our study examines how these two factors specifically affect GW separation.

\begin{figure*}[htbp]
        \includegraphics[scale=0.246]{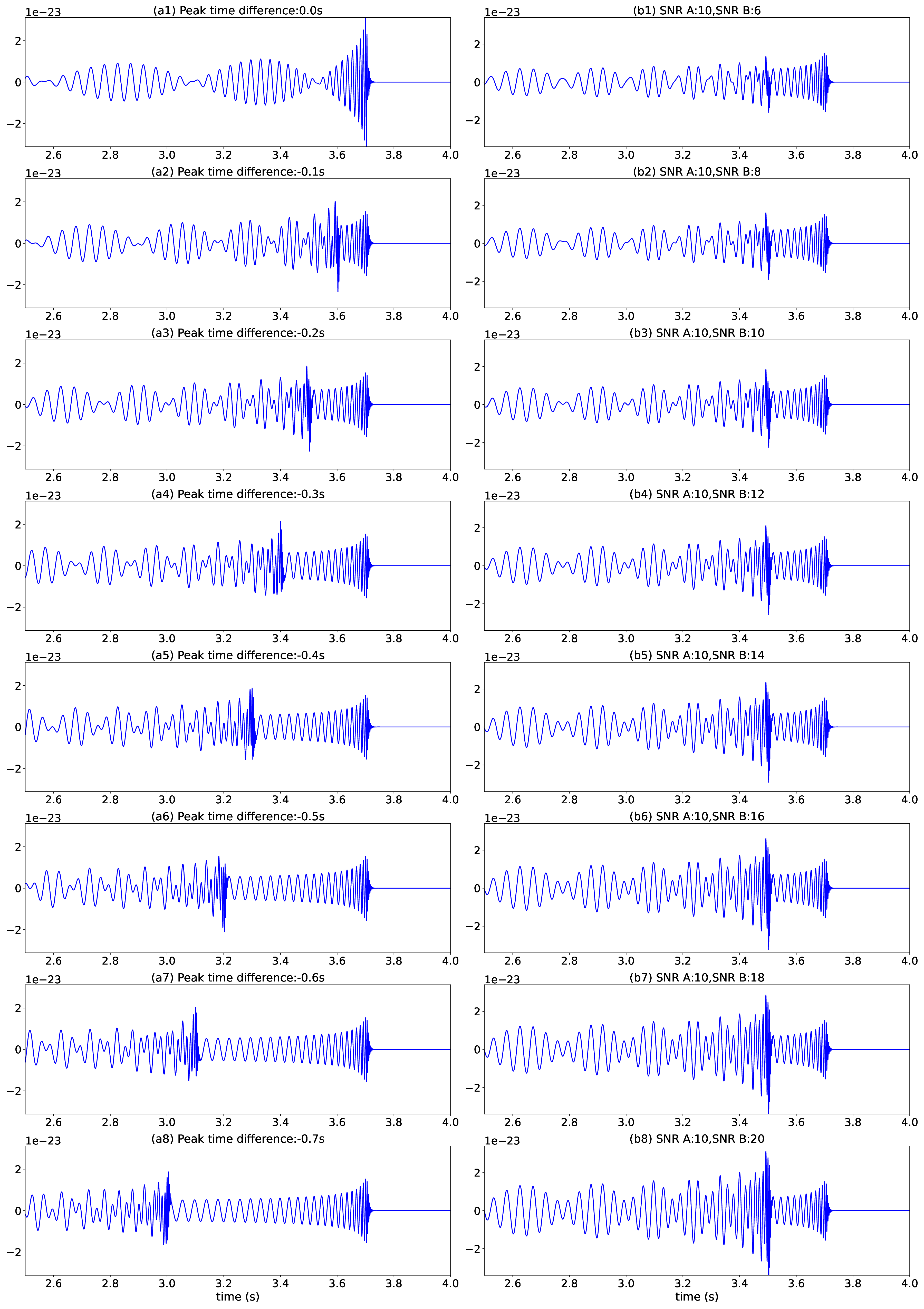}
        \caption{\label{fig:4}The overlapping signal waveforms in the time domain, excluding noise for clarity. Signal A originates from a source with parameters $Mass1=54\ M_\odot$, $Mass2=18\ M_\odot$, while Signal B corresponds to a source with $Mass1=32\ M_\odot$, $Mass2=20\ M_\odot$. Panel (a) demonstrates how the waveform changes as the peak time difference between Signal A and Signal B (specifically, $\mathit{peak\_time}(B) - \mathit{peak\_time}(A)$) varies. Panel (b) illustrates variations in the waveform according to differences in SNR, where the SNR of Signal A remains fixed at 10 while the SNR of Signal B fluctuates.
        }
\end{figure*}

Fig.~\ref{fig:4} illustrates an example of overlapping signal shapes, considering variations in peak time differences (a) and signal-to-noise ratio (SNR) differences (b). In subsequent sub-sections, we will introduce noise to these waveforms to produce simulated strain data, and then evaluate the performance of the GW separation model using this simulated data. From this figure, it is evident that, in most scenarios, the near merger and ringdown stages of signal A remain unaffected, whereas all stages of signal B appear blurred.

The following subsections will demonstrate that despite the blurring of signal B and the inspiral stage of signal A, in most cases, the waveforms of both signal A and signal B can often be accurately reconstructed.

\begin{figure*}[htbp]
        \includegraphics[scale=0.240]{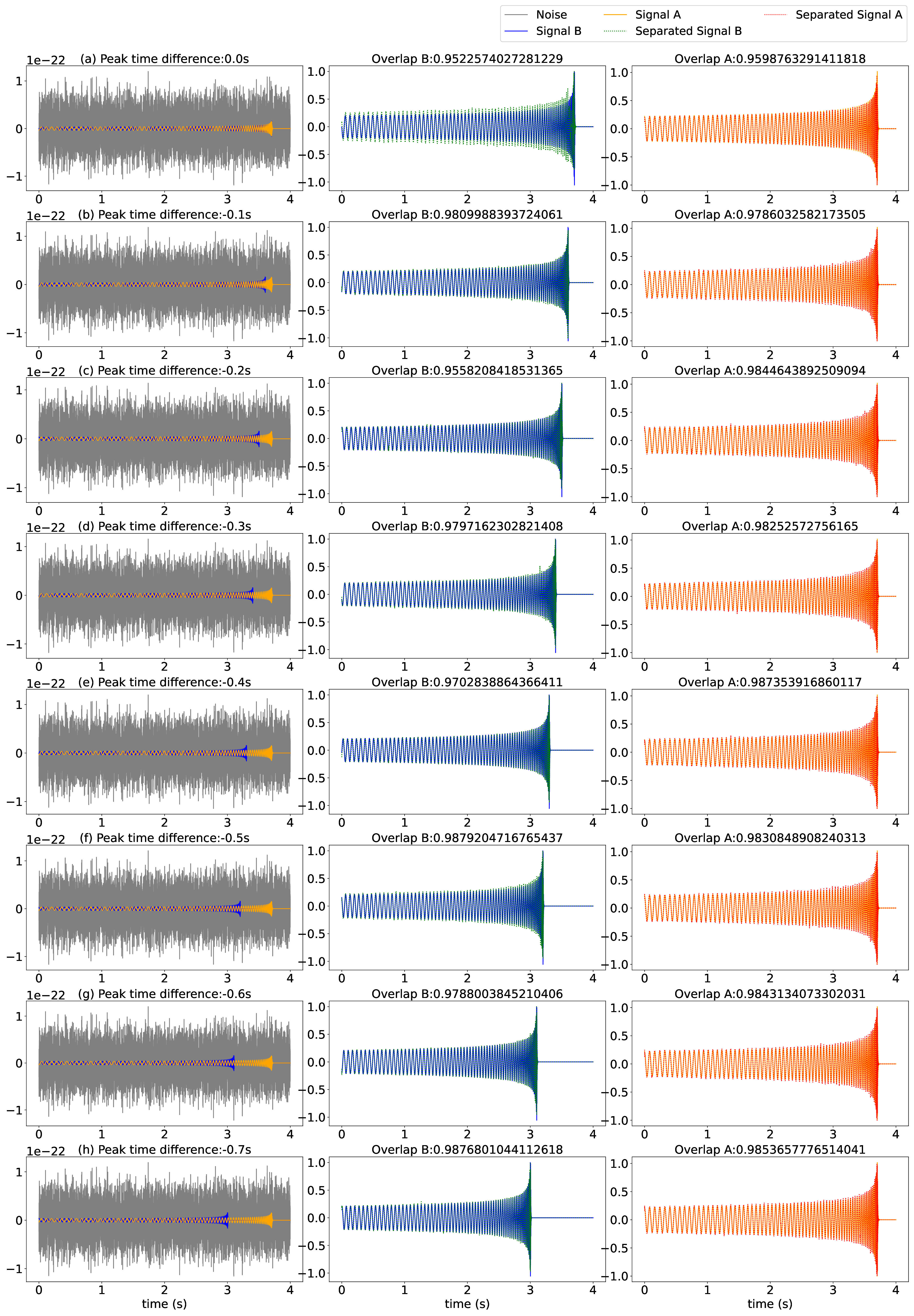}
        \caption{\label{fig:5}From top to bottom, the peak time difference between the two signals is incrementally set from 0 to 0.7 seconds in steps of 0.1 seconds. The leftmost column shows the mixed signal of Signal A, Signal B, and Noise. The right columns display the denoised and separated results, where the normalized separated signals are compared with the true source signals.}
\end{figure*}

\begin{figure*}[htbp]
        \includegraphics[scale=0.55]{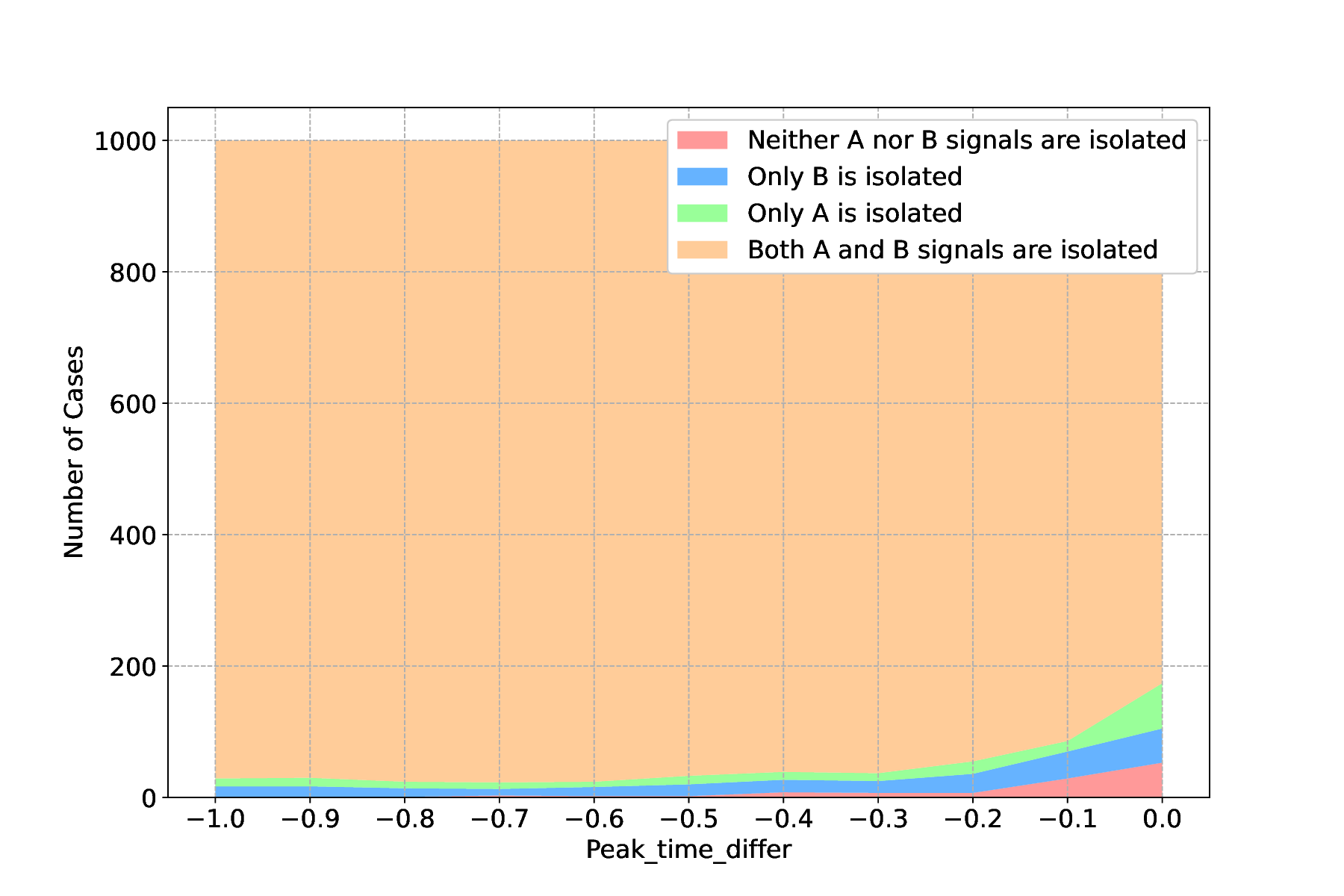}
        \caption{\label{fig:6}the separation outcomes for each peak time difference ranging from -1 to 0 second, increasing in increments of 0.1 s. The test dataset of each peak time difference has 1000 samples. The plot organizes the separation results into distinct categories, denoted by different colors: orange signifies successful separation of both signals (where the overlap between each separated signal and its corresponding source signal exceeds 0.9); green and blue represent cases in which either signal A or signal B, respectively, is successfully separated (where only one signal overlaps with its source by more than 0.9); and red indicates scenarios where neither signal is successfully separated (both signals overlap with their sources by less than 0.9).}
\end{figure*}

\begin{figure*}[htbp]
        \includegraphics[scale=0.236]{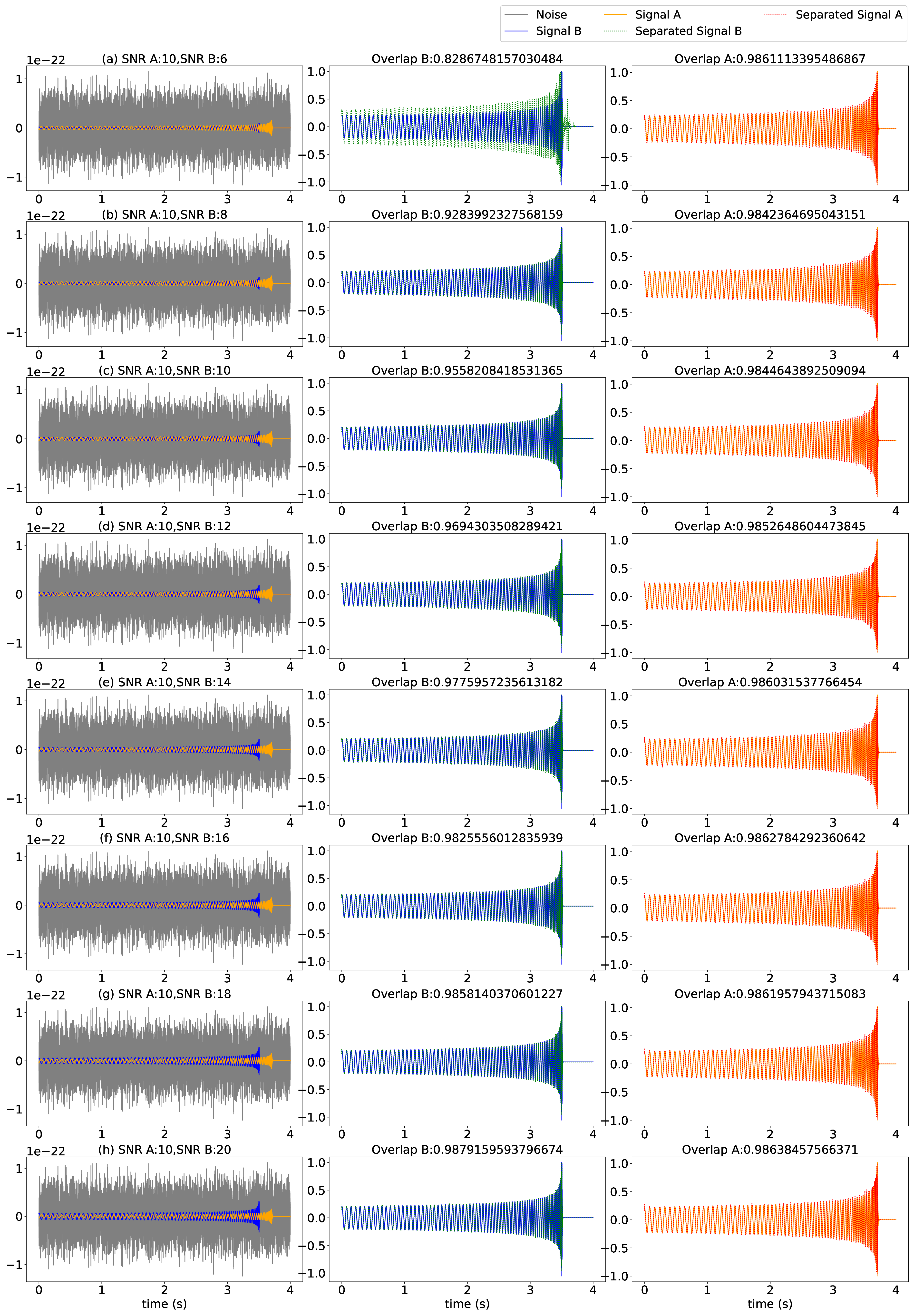}
        \caption{\label{fig:7}From top to bottom, the denoising and separation results for mixed signals are shown as the SNR difference between the two signals increases from -4 to 10 in step 2. The leftmost column displays the mixed signals of Signal A, Signal B, and Noise, while the columns to the right compare the two normalized separation results with their corresponding true source signals.
        }
\end{figure*}

\begin{figure*}[htbp]
        \includegraphics[scale=0.55]{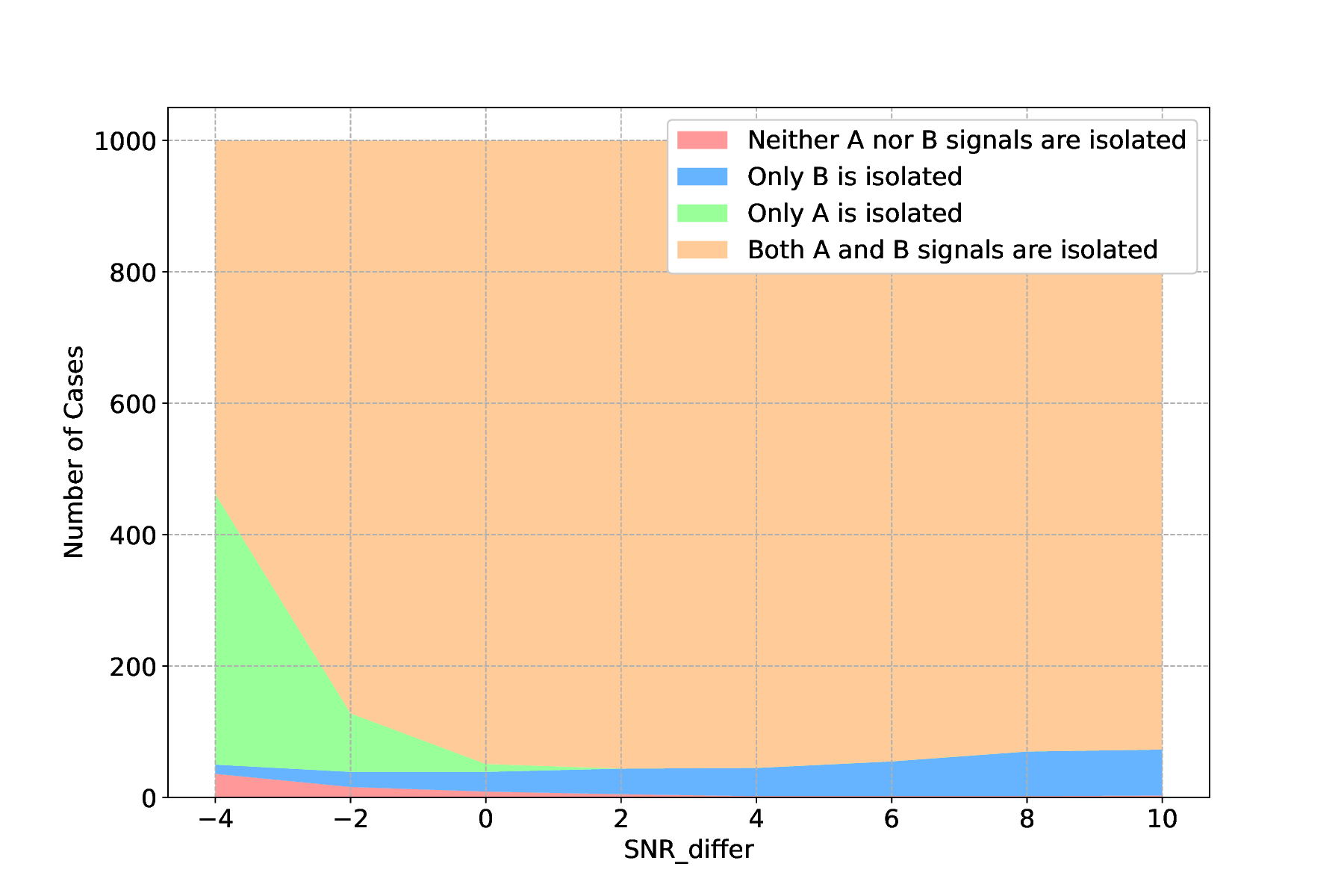}
        \caption{\label{fig:8}Separation outcomes for 1,000 samples generated for each SNR difference, ranging from -4 to 10 in steps of 2. The orange section represents scenarios where both signals are successfully separated (both signals have an overlap greater than 0.9 with their corresponding source signals). The green and blue sections represent cases where only Signal A or Signal B, respectively, is successfully separated (only one signal has an overlap greater than 0.9). The red section indicates scenarios where neither signal was successfully separated (both signals have an overlap of less than 0.9).}
\end{figure*}

\subsection{Impact of peak time difference on the GW separation\label{subsection:4-1}}

In this subsection, we elaborate on the influence of peak time disparities on GW separation. We produce three elements constituting a single strain: noise, signal A, and signal B. The source parameters of signal A and signal B are the same as the waveform shown in Fig.~\ref{fig:4}. With signal A peaking at 3.7 seconds within the entire strain window, we adjust the peak time of signal B to generate eight distinct waveforms. These waveforms exhibit time differences between the peaks of signal A and signal B ranging from -0.7 s to 0 s. By combining these three components, we synthesize eight unique strains. Subsequently, we subject these strains to the GW separation network and analyze the outputs. Fig.~\ref{fig:5} displays the individual outputs corresponding to each of the eight strains. To measure the separation performance, we utilize the overlap between the two separated signals and the two original signals. The overlap of signal $h$ and $\widetilde{h}$ can be written as

\begin{equation}
        overlap\big(h,\tilde{h}\big)=\frac{\int h(t)\tilde{h} (t)dt}{\sqrt{\int h^{2}(t)dt\int\tilde{h}^{2}(t) dt}} 
\end{equation}

\noindent From Fig.~\ref{fig:5} we can see that all eight strains have been successfully separated. Surprisingly, in extreme situations where the peak time of signal A and signal B are the same, the overlaps of both signal A and signal B are greater than 0.95.

% \begin{figure*}[htbp]
%         \includegraphics[scale=0.240]{peak_time_example}
%         \caption{\label{fig:5}From top to bottom, the peak time difference between the two signals is incrementally set from 0 to 0.7 seconds in steps of 0.1 seconds. The leftmost column shows the mixed signal of Signal A, Signal B, and Noise. The right columns display the denoised and separated results, where the normalized separated signals are compared with the true source signals.}
% \end{figure*}

Fig.~\ref{fig:5} presents a single case study demonstrating the effect of peak time difference on GW separation. Here, we undertake a comprehensive statistical analysis to investigate the broader influence of peak time disparities on the process of GW separation. To this end, we have generated eleven sub-test-datasets, with the sole difference among them being the peak time disparities, specifically \{-1.0 s, -0.9 s, -0.8 s, -0.7 s, -0.6 s, -0.5 s, -0.4 s, -0.3 s, -0.2 s, -0.1 s, 0 s\}. Each of these sub-datasets comprises 1000 samples, ensuring consistency in noise distribution and other parameter distribution across all datasets. The stack plot in Fig.~\ref{fig:6} illustrates the distribution of separated signals based on their relative merger time ($T_B - T_A$). Please note that if the overlap between the isolated signal and the actual injected signal exceeds 0.9, we consider the signal to be successfully isolated. This figure reveals that in most scenarios, both signal A and signal B are effectively separated. Notably, even in the most extreme circumstance, where the merger time of signal A and signal B coincide, over 80\% of the samples are still accurately separated, while approximately 10\% of the samples yield successful separation of only one of the two injections. Approximately 5\% of the samples show unsuccessful separation for both signal A and signal B. These results further underscore the exceptional performance of our model in denoising and separating mixed signals. The model effectively distinguishes overlapped signals under different peak time difference conditions, achieving high-quality separation results in the majority of cases. This highlights its robustness and capability in signal-processing tasks.

\subsection{Impact of SNR difference on the GW separation\label{subsection:4-2}}

In the preceding section, we discussed the impact of peak time differences on the separation of gravitational wave signals. In practical scenarios, the amplitudes of the individual components within the entangled signals exhibit diversity. Herein, we delve into the influence of signal strength on GW disentanglement. Signal strength can be quantified by the matched signal-to-noise ratio (SNR). To be specific, we maintain an SNR of 10 for signal A while adjusting the SNR differential between signal B and signal A in increments of 2, spanning from -4 to 10. Consequently, the SNRs for signal B are adjusted to the following values: \{6, 8, 10, 12, 14, 16, 18, 20\}.

We configure the parameters identically to those presented in Fig.~\ref{fig:4}. Specifically, we establish the peak time of signal A at 3.7 seconds within the strain window and set the peak time of signal B at 3.5 seconds, resulting in a peak time difference of -0.2 seconds. We then adjusted the SNR of signal B, varying it from 6 to 20. After superimposing signal A, signal B, and noise, we input the combined signal into the Gravitational Wave (GW) separation network and obtained the output. Fig.~\ref{fig:7} illustrates the separated and injected waveforms for both signal A and signal B.

Here, we analyze the influence of signal A on the GW separation performance of signal B by the right column of Fig.~\ref{fig:7}. By changing the SNR of signal B from 6 to 20, the separation results of signal A almost unchanged. All the separation overlaps of signal A are greater than 0.98.

When the Signal-to-Noise Ratio (SNR) of signal B is 6, we can see that the overlap between the separated signal B and the buried signal is approximately 0.82. We hypothesize that there may be two primary factors influencing the separation performance of signal B. Firstly, the SNR of signal B is significantly low, causing noise to interfere with the separation process. Secondly, both signal A and noise contribute to the decrease in separation performance. To gain a deeper understanding of the reasons behind the incorrect separation, we subtract signal A and preserve only signal B and the noise in the strain data. This modified data is then inputted into the separation model to observe the impact on the separation of signal B. We verified that the overlap of signal B is equal to 0.80, which is nearly identical to 0.82. The results suggest that the underwhelming performance observed in the separation of signal B in Fig.~\ref{fig:7} (a) is unrelated to the overlapping signal B, but is instead impacted by the intensity of noise.

% \begin{figure*}[htbp]
%         \includegraphics[scale=0.236]{snr_example}
%         \caption{\label{fig:7}From top to bottom, the denoising and separation results for mixed signals are shown as the SNR difference between the two signals increases from -4 to 10 in step 2. The leftmost column displays the mixed signals of Signal A + Signal B + Noise, while the columns to the right compare the two normalized separation results with their corresponding true source signals.
%         }
% \end{figure*}
 
% \begin{figure*}[htbp]
%         \includegraphics[scale=0.55]{snr_test}
%         \caption{\label{fig:8}Separation outcomes for 1,000 samples generated for each SNR difference, ranging from -4 to 10 in steps of 2. The orange section represents scenarios where both signals are successfully separated (both signals have an overlap greater than 0.9 with their corresponding source signals). The green and blue sections represent cases where only Signal A or Signal B, respectively, is successfully separated (only one signal has an overlap greater than 0.9). The red section indicates scenarios where neither signal was successfully separated (both signals have an overlap of less than 0.9).}
% \end{figure*}

To further investigate the impact of SNR differences on separation performance and identify potential shortcomings of our model, we prepared 1,000 samples for each SNR difference value. Fig.~\ref{fig:8} illustrates the four separation scenarios under different SNR differences, with the x-axis representing SNR differences ranging from -4 to 10. Note that the SNR of signal A is set to 10. We set the SNR of signal B to \{6, 8, 10, 12, 14, 16, 18, 20\} corresponding to the SNR difference \{-4, -2, 0, 2, 4, 6, 8, 10\}. From the area chart in Figure 8, it is evident that the orange region, indicating the successful separation of both signals, occupies the majority of the area. The red region, representing scenarios where neither signal was successfully separated, remains very small. Specifically, when the SNR of Signal A is fixed at 10 and the SNR of Signal B is 6 or 8, the instances where only Signal A is successfully separated significantly outnumber the instances where only Signal B is successfully separated. This indicates that, in scenarios with smaller SNR differences, the model is more likely to successfully separate the signal with the higher SNR. These results suggest that further optimization is needed to enhance the model's performance in separating overlapping signals with low SNR parts. At the same time, they also confirm the robustness of the current model in most cases.

\section{Conclusion\label{section:5}}

In this paper, we attempt to address the challenge posed by overlapping GW signals, which is an emerging issue as future GW observatories. We have demonstrated the feasibility of adapting speech separation techniques to the domain of GW signal separation, employing deep learning models for this task. Our findings reveal that the proposed approach can effectively disentangle overlapping GW signals, even when they exhibit different peak time differences. This capability ensures robust signal identification and accurate extraction of individual GW events from a complex signal mixture. Additionally, we observed that the method performs remarkably well across a range of SNRs. Even in low SNR scenarios, where noise levels are relatively high, the model demonstrates its ability to separate and identify GW signals with reasonable accuracy.

\nocite{*}

\bibliography{apssamp}% Produces the bibliography via BibTeX.

\end{document}